\documentclass[aps,pra,superscriptaddress,singlecolumn,showpacs]{revtex4}
\usepackage{amsfonts}
\usepackage{amsmath}
\usepackage{mathrsfs}
\usepackage{amssymb}
\usepackage{graphicx}

\begin{document}

\title{Ground-state properties of spin-orbit-coupled dipolar Bose-Einstein
condensates with in-plane gradient magnetic field}
\author{Xiaoqian Li}
\affiliation{Key Laboratory for Microstructural Material Physics of Hebei Province,
School of Science, Yanshan University, Qinhuangdao 066004, China}
\author{Qingbo Wang}
\affiliation{Key Laboratory for Microstructural Material Physics of Hebei Province,
School of Science, Yanshan University, Qinhuangdao 066004, China}
\affiliation{Department of Physics, Tangshan Normal University, Tangshan 063000, China}
\author{Huan Wang}
\affiliation{Key Laboratory for Microstructural Material Physics of Hebei Province,
School of Science, Yanshan University, Qinhuangdao 066004, China}
\affiliation{Department of Applied Physics, School of Science, Xi'an Jiaotong University,
Xi'an 710049, Shaanxi, China}
\author{Chunxiao Shi}
\affiliation{Key Laboratory for Microstructural Material Physics of Hebei Province,
School of Science, Yanshan University, Qinhuangdao 066004, China}
\author{Malcolm Jardine}
\affiliation{Department of Physics and Astronomy, University of Pittsburgh, Pittsburgh,
Pennsylvania 15260, USA}
\author{Linghua Wen}
\email{linghuawen@ysu.edu.cn}
\affiliation{Key Laboratory for Microstructural Material Physics of Hebei Province,
School of Science, Yanshan University, Qinhuangdao 066004, China}
\affiliation{Department of Physics and Astronomy, University of Pittsburgh, Pittsburgh,
Pennsylvania 15260, USA}
\date{\today}

\begin{abstract}
We investigate the ground-state properties of spin-orbit-coupled
pseudo-spin-1/2 dipolar Bose-Einstein condensates (BECs) in a
two-dimensional harmonic trap and an in-plane quadrupole field. The effects
of spin-orbit coupling (SOC), dipole-dipole interaction (DDI) and the
in-plane quadrupole field on the ground-state structures and spin textures
of the system are systematically analyzed and discussed. For fixed SOC and
DDI strengths, the system shows a quadrupole stripe phase with a
half-quantum vortex, or a quadrupole Thomas-Fermi phase with a half-quantum
antivortex for small quadrupole field strength, depending on the ratio
between inter- and intraspecies interaction. As the quadrupole field
strength enhances, the system realizes a ring mixed phase with a hidden
vortex-antivortex cluster rather than an ordinary giant vortex in each
component. Of particular interest, when the strengths of DDI and quadrupole
field are fixed, strong SOC leads to the formation of criss-crossed vortex
string structure. For given SOC and quadrupole field, the system for strong
DDI displays a sandwich-like structure, or a special delaminated structure
with a prolate antivortex in the spin-up component. In addition, typical
spin textures for the ground states of the system are analyzed. It is shown
that the system sustains exotic topological structures, such as a hyperbolic
spin domain wall, skyrmion-half-antiskyrmion-antiskyrmion lattice,
half-skyrmion-skyrmion-half-antiskyrmion lattice, and a drum-shaped
antimeron.
\end{abstract}

\maketitle

\section{Introduction}

One of the most fascinating recent developments in physics has been the
production of dipolar quantum gases with dipole-dipole interaction (DDI)
\cite{Lahaye1,Ray1,Ray2}. When the electric or magnetic DDI of the ultracold
quantum gas cannot be neglected, we must consider the dipolar effects
between atoms or molecules. Essentially, the DDI is long-range and
anisotropic, which has important influences on the static structures,
dynamic properties and stability of ultracold quantum gases \cite%
{Goral,Yi,Santos,Malet,Kadau,Cha,Zou,Oshima,Singh,Xi}. Relevant studies show
that DDI can lead to more novel, fascinating and even unexpected effects
than the conventional contact $s$-wave interaction in quantum gases \cite%
{Lahaye1,Ray1,Ray2,Chomaz}. In addition, the anisotropic interaction between
dipoles gives us more controllable physical quantities, which makes dipolar
quantum gases more likely to be used in many potential application based
frontier fields such as quantum simulation and quantum computation.
Recently, the long-range and anisotropic magnetic DDI has been
experimentally observed in Bose-Einstein condensates (BECs) with $^{52}$Cr, $%
^{164}$Dy and $^{168}$Er atoms \cite{Chomaz,Lahaye2,Lu,Aikawa,Lepoutre},
which provides an opportunity to test and further develop the current
theories on cold atom physics.

In addition, the spin-orbit-coupled quantum gases have also become one of
the frontier research fields in physics in recent years \cite%
{Dalibard,Lin,Cheuk,Meng,Wu,Zhai}. Numerous experimental and theoretical
studies have shown that spin-orbit coupling (SOC) in cold atom gases can
lead to many novel quantum phases that have rich physical properties, such
as stripe phase, topological superfluid phase, half-quantum vortex, soliton
excitation, Zitterbewegung oscillation, and collective modes \cite%
{Wu,Zhai,Sinha,Ramachandhran,YZhang,Xu,Qu,YLi}. It seems obvious that there
is particular interest in investigating the combined effects of DDI and SOC
on spinor BECs, and this idea has recently attracted considerable attention
\cite{Deng,Wilson,Gopalakrishnan,Kato}.

In this paper, we study the ground-state structures and spin textures of
spin-orbit-coupled dipolar BECs in a harmonic trap and an in-plane gradient
magnetic field. Combined effects of DDI, SOC, and the gradient magnetic
field on the ground-state properties of the system are analyzed. We find
that the system supports a quadrupole stripe phase with a half-quantum
vortex and a quadrupole Thomas-Fermi (TF) phase with a half-quantum
antivortex for small strength of in-plane quadrupole field (i.e. in-plane
gradient magnetic field). With the increase of quadrupole field strength, a
ring mixed phase with a hidden vortex-antivortex lattice cluster \cite%
{Wen1,Mithun,Wen2} is formed in each component. For given strengths of DDI
and quadrupole field, strong SOC results in the generation of criss-crossed
vortex strings. For fixed gradient magnetic field and strong SOC, the ground
state exhibits a sandwich-like structure or a delaminated half-quantum
antivortex structure. Furthermore, the system displays rich spin textures
and topological structures, such as a hyperbolic domain wall,
skyrmion-half-antiskyrmion-antiskyrmion lattice,
half-skyrmion-skyrmion-half-antiskyrmion lattice, and drum-shaped
half-antiskyrmion (antimeron).

The paper is organized as follows. In Sec. II, we introduce the theoretical
model for the system. In Sec. III, we present and analyze the ground-state
structures and typical spin textures of the system. Our main findings are
summarized in Sec. IV.

\section{Formalism}

We consider a quasi-two-dimensional (quasi-2D) system of Rashba-type
spin-orbit-coupled dipolar BECs in a harmonic trap and an in-plane gradient
magnetic field \cite{Ray1,Lin,Zhai,Zhang1}. The magnetic dipoles are fully
polarized by an auxiliary magnetic filed which is in the $x$-$z$ plane and
forms an angle $\alpha $ with the $z$-axis \cite{Malet,Zhang2}. The dynamics
of the system obeys the generalized coupled Gross-Pitaevskii (GP) equations%
\begin{eqnarray}
i\hbar \frac{\partial \psi _{1}}{\partial t} &=&\left[ -\frac{\hbar
^{2}\nabla ^{2}}{2m}+V(\mathbf{r})+g_{11}\left\vert \psi _{1}\right\vert
^{2}+g_{12}\left\vert \psi _{2}\right\vert ^{2}\right] \psi _{1}+\left(
C_{dd}^{11}\int U_{dd}\left\vert \psi _{1}(\mathbf{r}^{\prime
},t)\right\vert ^{2}d\mathbf{r}^{\prime }+C_{dd}^{12}\int U_{dd}\left\vert
\psi _{2}(\mathbf{r}^{\prime },t)\right\vert ^{2}d\mathbf{r}^{\prime
}\right) \psi _{1}  \notag \\
&&+\hbar \left( \lambda _{x}\partial _{x}-i\lambda _{y}\partial _{y}\right)
\psi _{2}+g_{F}\mu _{B}B(x+iy)\psi _{2},  \label{GP1} \\
i\hbar \frac{\partial \psi _{2}}{\partial t} &=&\left[ -\frac{\hbar
^{2}\nabla ^{2}}{2m}+V(\mathbf{r})+g_{21}\left\vert \psi _{1}\right\vert
^{2}+g_{22}\left\vert \psi _{2}\right\vert ^{2}\right] \psi _{2}+\left(
C_{dd}^{22}\int U_{dd}\left\vert \psi _{2}(\mathbf{r}^{\prime
},t)\right\vert ^{2}d\mathbf{r}^{\prime }+C_{dd}^{21}\int U_{dd}\left\vert
\psi _{1}(\mathbf{r}^{\prime },t)\right\vert ^{2}d\mathbf{r}^{\prime
}\right) \psi _{2}  \notag \\
&&-\hbar \left( \lambda _{x}\partial _{x}+i\lambda _{y}\partial _{y}\right)
\psi _{1}+g_{F}\mu _{B}B(x-iy)\psi _{1},  \label{GP2}
\end{eqnarray}%
where $\psi _{j}$ $(j=1,2)$ is the component wave function, with 1 and 2
corresponding to spin-up and spin-down, respectively. We assume that the two
component atoms have the same mass $m$. The coefficients $g_{jj}=2\sqrt{2\pi
}a_{j}\hbar ^{2}/ma_{z}$ $(j=1,2)$, $g_{12}=g_{21}=\sqrt{2\pi }a_{12}\hbar
^{2}/ma_{z}$ describe the intra- and interspecies interaction strengths,
which are directly related to the $s$-wave scattering lengths $a_{j}$ and $%
a_{12}$ between intra- and intercomponent atoms, and $a_{z}=\sqrt{\hbar
/m\omega _{z}}$ is the oscillation length in the $z$ direction. $V(\mathbf{r}%
)=m\omega _{\perp }^{2}(x^{2}+y^{2})/2$ is the external trapping potential,
with $\omega _{\perp }=$ $\omega _{x}=$ $\omega _{y}$ being the harmonic
trap frequency and $r=\sqrt{x^{2}+y^{2}}$. Here $\lambda _{x}$ and $\lambda
_{y}$ are the SOC strengths in the $x$- and $y$- directions \cite{Lin}. $%
g_{F}=-1/2$ is the Lande factor, $\mu _{B}$ is Bohr magnetic moment, and $B$
denotes the strength of in-plane quadrupole magnetic field \cite%
{Leanhardt,Ray1,Ray2}. $C_{dd}^{jj}=\mu _{0}\mu _{j}^{2}/4\pi $ $(j=1,2)$
and $C_{dd}^{12}=C_{dd}^{21}=\mu _{0}\mu _{1}\mu _{2}/4\pi $ are the
magnetic DDI constants of intraspecies and interspecies, respectively, where
$\mu _{0}$ is the magnetic permeability of vacuum, and $\mu _{j}$ $(j=1,2)$
represents magnetic dipole moment of the $j$-th component atom. We assume
that $\mu _{1}=\mu _{2}=\mu $, which means $%
C_{dd}^{11}=C_{dd}^{22}=C_{dd}^{12}=C_{dd}^{21}=C_{dd}$ \cite{Xu2}. The
long-range and nonlocal DDI can be expressed as \cite{Adhikari}%
\begin{equation}
U_{dd}(\mathbf{r-r}^{\prime })=\frac{1-3\cos ^{2}\theta }{\left\vert \mathbf{%
r-r}^{\prime }\right\vert ^{3}},  \label{DDI}
\end{equation}%
where $\theta $ is the angle between the polarization direction and the
relative position of the atoms. The normalization condition of the system is
given by$\int [\left\vert \psi _{1}\right\vert ^{2}+\left\vert \psi
_{2}\right\vert ^{2}]dxdy=N$, with $N$ being the number of atoms.

For the convenience of numerical calculation, we introduce the dimensionless
parameters via the notations $r^{\prime }=r/a_{0}$, $a_{0}=\sqrt{\hbar
/m\omega _{\perp }}$, $t^{\prime }=\omega _{\perp }t$, $V^{\prime
}(r)=V(r)/\hbar \omega _{\perp }$, $\psi _{j}^{\prime }=\psi _{j}a_{0}/\sqrt{%
N}(j=1,2)$ and $B^{\prime }=g_{F}\mu _{B}a_{0}B/\hbar \omega _{\perp }$, and
then we obtain the dimensionless GP equations%
\begin{eqnarray}
i\partial _{t}\psi _{1} &=&\left( -\frac{1}{2}\nabla ^{2}+V+\beta
_{11}\left\vert \psi _{1}\right\vert ^{2}+\beta _{12}\left\vert \psi
_{2}\right\vert ^{2}\right) \psi _{1}+\gamma _{dd}\mathscr{F}_{2D}^{-1}[%
\widetilde{n}(\mathbf{k},t)F(\mathbf{k}a_{z}/\sqrt{2})]\psi _{1}  \notag \\
&&+\left( \lambda _{x}\partial x-i\lambda _{y}\partial y\right) \psi
_{2}+B(x+iy)\psi _{2},  \label{DimensionlessGP1} \\
i\partial _{t}\psi _{2} &=&\left( -\frac{1}{2}\nabla ^{2}+V+\beta
_{12}\left\vert \psi _{1}\right\vert ^{2}+\beta _{22}\left\vert \psi
_{2}\right\vert ^{2}\right) \psi _{2}+\gamma _{dd}\mathscr{F}_{2D}^{-1}[%
\widetilde{n}(\mathbf{k},t)F(\mathbf{k}a_{z}/\sqrt{2})]\psi _{2}  \notag \\
&&-\left( \lambda _{x}\partial x+i\lambda _{y}\partial y\right) \psi
_{1}+B(x-iy)\psi _{1},  \label{DimensionlessGP2}
\end{eqnarray}%
where the prime is omitted for brevity. Here $\beta _{jj}=2\sqrt{2\pi }%
a_{j}N/a_{z}$ $(j=1,2)$ and $\beta _{12}=\beta _{21}=\sqrt{2\pi }%
a_{12}N/a_{z}$ are the dimensionless intra- and interspecies interaction
strengths. The dipolar coupling constant is given by $\gamma _{dd}=\mu
_{0}\mu ^{2}mN/(3\sqrt{2\pi }\hbar ^{2}a_{z})$. $\mathscr{F}_{2D}$ is the
two-dimensional Fourier transform operator, and $\widetilde{n}$($\mathbf{k}$,%
$t$)$=\mathscr{F}_{2D}[n(\mathbf{r},t)]$ \cite{Zhang2}. The function $F$($%
\mathbf{q}$) with $\mathbf{q}\equiv \mathbf{k}a_{z}/\sqrt{2}$ denotes the $k$
space DDI for the quasi-$2D$ geometry, which is composed of two parts,
originating from polarization perpendicular or parallel to the direction of
the dipole tilt. Specifically, $F$($\mathbf{q}$)$=$cos$^{2}$($\alpha $)$%
F_{\bot }$($\mathbf{q}$)$+$sin$^{2}$($\alpha $)$F_{\Vert }$($\mathbf{q}$),
where $\alpha $ is the angle between the $z$-axis and the polarization
vector $\hat{d}$, $F_{\perp }$($\mathbf{q}$)$=2-3\sqrt{\pi }qe^{q^{2}}$erfc$%
(q)$, $F_{_{\Vert }}$($\mathbf{q}$)$=-1+3\sqrt{\pi }(q_{d}^{2}/q)e^{q^{2}}$%
erfc$(q)$, $\mathbf{q}_{d}$ is the wave vector along the direction of the
projection of $\hat{d}$ onto the $x$-$y$ plane, and erfc is the
complementary error function \cite{Fischer,Nath}. If the polarization is
perpendicular to the condensate plane, i.e., $\alpha =0$, one can get $F$($%
\mathbf{q}$)$=2-3\sqrt{\pi }qe^{q^{2}}$erfc$(q)$, which has been discussed
in previous work \cite{Shirley}.

In order to further understand the topological properties of the system,\ we
use a nonlinear Sigma model \cite{Kasamatsu,Wang} and introduce a normalized
complex-valued spinor $\chi =\left[ \chi _{1},\chi _{2}\right] ^{T}$ with
the normalization condition $\left\vert \chi _{1}\right\vert ^{2}+\left\vert
\chi _{2}\right\vert ^{2}=1$. The component wave function is $\psi _{j}=%
\sqrt{\rho }\chi _{j}$ $\left( j=1,2\right) $, and the total density of the
system is $\rho =\left\vert \psi _{1}\right\vert ^{2}+\left\vert \psi
_{2}\right\vert ^{2}$. The spin density is expressed by $\mathbf{S=}%
\overline{\chi }\mathbf{\sigma \chi }$\textbf{,} where $\sigma =\left(
\sigma _{x},\sigma _{y},\sigma _{z}\right) $ are the Pauli matrices. The
components of $\mathbf{S}$ can be written as%
\begin{eqnarray}
S_{x} &=&\left( \psi _{1}^{\ast }\psi _{2}+\psi _{2}^{\ast }\psi _{1}\right)
/\rho ,  \label{SX} \\
S_{y} &=&-i\left( \psi _{1}^{\ast }\psi _{2}-\psi _{2}^{\ast }\psi
_{1}\right) /\rho ,  \label{SY} \\
S_{z} &=&\left( \left\vert \psi _{1}\right\vert ^{2}-\left\vert \psi
_{2}\right\vert ^{2}\right) /\rho ,  \label{SZ}
\end{eqnarray}%
with $\left\vert \mathbf{S}\right\vert =\sqrt{S_{x}^{2}+S_{y}^{2}+S_{z}^{2}}%
=1$. The spacial distribution of the topological structure of the system is
described by the topological charge density%
\begin{equation}
q(r)=\frac{1}{4\pi }\mathbf{S}\bullet \left( \frac{\partial \mathbf{S}}{%
\partial x}\times \frac{\partial \mathbf{S}}{\partial y}\right) ,
\label{TopologicalChargeDensity}
\end{equation}%
and the topological charge $Q$ is defined as%
\begin{equation}
Q=\int q(r)dxdy.  \label{TopologicalCharge}
\end{equation}

\section{Ground-state structures and spin textures}

Here the system is rather complex. To the best of our knowledge, there is no
analytical solution for this system. In the following, we numerically solve
the GP equations (\ref{DimensionlessGP1}) and (\ref{DimensionlessGP2}) and
obtain the ground state of the system by using the imaginary-time
propagation method \cite{YZhang,Xu,Wen3}. For clarity, we say that there is
initial phase separation (i.e. component separation) when the contact
interaction parameters are chosen to be $\beta _{_{11}}=\beta _{_{22}}=800$
and $\beta _{_{12}}=1600$. And we say there is initial phase mixing (i.e.
component mixing) when the contact interaction parameters are chosen to be $%
\beta _{_{11}}=\beta _{_{22}}=800$ and $\beta _{_{12}}=600$. These
parameters are essentially in agreement with the relevant parameters in
physical experiment of BECs. In addition, an $\alpha =\pi /2$ is considered
in the present work. Our results show that the system can exhibit rich and
exotic ground-state structures and spin textures, which we will now discuss
by considering how each parameter can affect the ground state.

\subsection{Role of in-plane quadrupole field}

We first study the effect of the in-plane gradient magnetic field on the
ground-state properties of the system with fixed SOC and DDI. Figure 1 shows
the density distributions (odd rows) and the corresponding phase
distributions (even rows) for the ground states of the system. In these
calculations the parameters are given by $\gamma _{dd}=0.5$, $\lambda
_{x}=\lambda _{y}=2$, $\beta _{11}=\beta _{22}=800$, then $\beta _{12}=1600$
for the left four columns and $\beta _{12}=600$ for the right three columns,
and the strengths of the in-plane gradient magnetic field are given in
figure 1. Here the top two rows and the bottom two rows denote component 1
(spin-up component) and component 2 (spin-down component), respectively.

\begin{figure*}[tbph]
\centerline{\includegraphics*[width=10 cm]{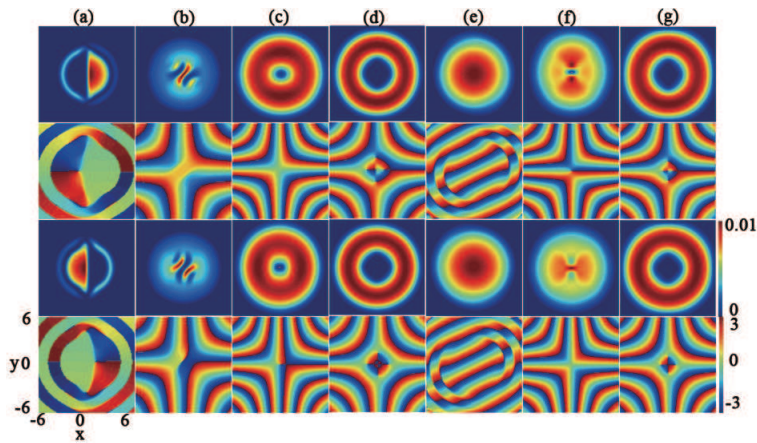}}
\caption{(Color online) Ground states of spin-orbit-coupled dipolar BECs
with in-plane quadrupole field in a harmonic trap. The rows from top to
bottom denote $\left\vert \protect\psi _{1}\right\vert ^{2}$, arg$\protect%
\psi _{1}$, $\left\vert \protect\psi _{2}\right\vert ^{2}$, and arg$\protect%
\psi _{2}$. The relevant parameters are $\protect\lambda _{x}=\protect%
\lambda _{y}=2$, $\protect\gamma _{dd}=0.5$, $\protect\beta _{11}=\protect%
\beta _{22}=800$, with $\protect\beta _{12}=1600$ (the left four columns)
and $\protect\beta _{_{12}}=600$ (the right three columns). The strengths of
the in-plane gradient magnetic field are given by (a) $B=0$, (b) $B=1.5$,
(c) $B=4$, (d) $B=6$, (e) $B=0$, (f) $B=1.5$, and (g) $B=5$. The unit length
is $a_{0}$.}
\label{Figure1}
\end{figure*}

First focus on figure 1(a), i.e. the first column, where this is the
situation that there is initial component separation and no applied gradient
magnetic field. The first and third row show that the two components have
immiscible combination density patterns with semicircle and crescent shapes,
and rows two and four display a hidden vortex (clockwise rotation) \cite%
{Wen1,Mithun,Wen2} in each component due to the presence of SOC and DDI.
Once weak quadrupole field is included, e.g. $B=1.5$ as in column (b), the
system exhibits an unusual, staggered stripe phase with a half-quantum
vortex, where the two component densities are spatially separated from each
other. The phase profiles show that there is a singly quantized hidden
vortex in the spin-down component and both the phase profiles display
quadrupole-like distribution \cite{JLi}. We call it a quadrupole stripe
phase with a half-quantum vortex to distinguish it from the conventional
stripe phase in spin-orbit-coupled spin-$1/2$ BECs \cite%
{Zhai,YZhang,CWang,XXu}. When the quadrupole field strength increases to $%
B=4 $, as in (c), the system displays obvious phase mixing with a singly
quantized visible vortex in the spin-down component and a dark soliton in
the spin-up component, as can be clearly seen in the figure. With the
further increase of quadrupole field strength, the density distributions of
the two components become almost the same with a large density hole in each
component (figure 1(d)). Here the density hole is not an ordinary giant
vortex but an exotic hidden vortex-antivortex cluster composed of several
hidden vortices and antivortices (anticlockwise rotation), which is quite
different from the previous cases of rotating BECs with or without SOC \cite%
{XXu,Fetter,XFZhou,Aftalion}.

For the case of initial phase mixing without the quadrupole filed, the
ground state is a typical plane-wave phase (i.e.TF phase) as displayed in
figure 1(e).\ When $B=1.5$, a singly quantized antivortex forms in the
spin-up component while the density of the spin-down component keeps a TF
distribution with the phase profile exhibiting quadrupole-like distribution
(figure 1(f)). We call it quadrupole TF phase with a half-quantum antivortex
to distinguish from the usual plane-wave phase. As the quadrupole field
strength increases to $B=5$ (figure 1(g)), the densities of the two
components are almost completely mixed with different hidden
vortex-antivortex clusters being generated in the central regions, which is
similar to that in figure 1(d).

\subsection{Role of SOC}

Next we consider the effects of SOC on the density and phase distributions
of the ground states of spin-orbit-coupled BECs, now with fixed DDI and
fixed in-plane gradient magnetic field, where the distributions are shown in
figure 2. The gradient magnetic field is $B=2.5$, the dipolar coupling is $%
\gamma _{dd}=0.5$, and the intra- and interspecies interaction strengths are
$\beta _{11}=\beta _{22}=800$, with $\beta _{12}=1600$ for the left three
columns and $\beta _{12}=600$ for the right three columns. The rows from top
to bottom represent $\left\vert \psi _{1}\right\vert ^{2}$, arg$\psi _{1}$, $%
\left\vert \psi _{2}\right\vert ^{2}$, and arg$\psi _{2}$, just as before.
The changing SOC strengths are $\lambda _{x}=\lambda _{y}=0$ (a), $\lambda
_{x}=\lambda _{y}=3$ (b), $\lambda _{x}=\lambda _{y}=5$ (c), $\lambda
_{x}=\lambda _{y}=1$ (d), $\lambda _{x}=\lambda _{y}=5$ (e), and $\lambda
_{x}=\lambda _{y}=10$ (f).

\begin{figure*}[tbph]
\centerline{\includegraphics*[width=10cm]{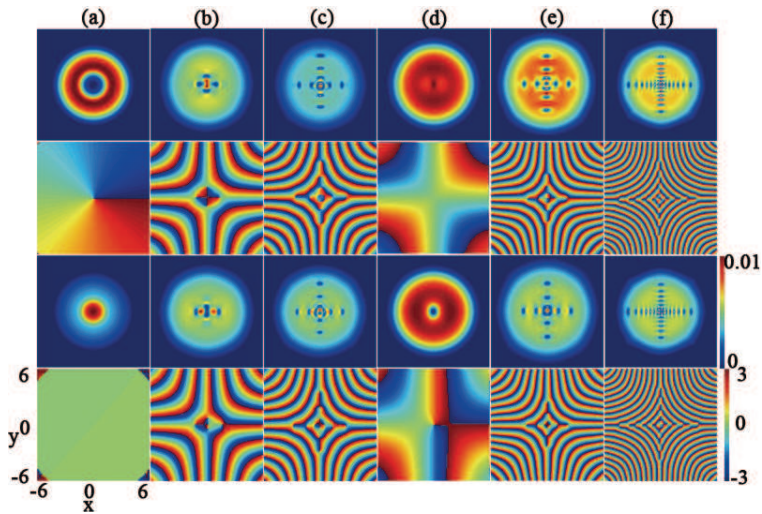}}
\caption{(Color online) Ground-state density distributions and phase
distributions for spin-orbit-coupled dipolar BECs in a harmonic trap and an
in-plane quadrupole field. The rows from top to bottom denote $\left\vert
\protect\psi _{1}\right\vert ^{2}$, arg$\protect\psi _{1}$, $\left\vert
\protect\psi _{2}\right\vert ^{2}$, and arg$\protect\psi _{2}$. The
parameters are $B=2.5$, $\protect\gamma _{dd}=0.5$, $\protect\beta _{11}=%
\protect\beta _{22}=800$, $\protect\beta _{12}=1600$ (left three columns),
and $\protect\beta _{12}=600$ (right three columns). (a) $\protect\lambda %
_{x}=\protect\lambda _{y}=0$, (b) $\protect\lambda _{x}=\protect\lambda %
_{y}=3$, (c) $\protect\lambda _{x}=\protect\lambda _{y}=5$, (d) $\protect%
\lambda _{x}=\protect\lambda _{y}=1$, (e) $\protect\lambda _{x}=\protect%
\lambda _{y}=5$, and (f) $\protect\lambda _{x}=\protect\lambda _{y}=10$. The
unit length is $a_{0}$.}
\label{Figure2}
\end{figure*}

For the case of initial component separation, when there is no SOC (i.e. $%
\lambda _{x}=\lambda _{y}=0$), the topological structure of the system is a
typical Anderson-Toulouse coreless antivortex \cite{Anderson}, where the
core of the circulating external component is filled with the other
nonrotating component (see figure 2(a)).\ With the increase of SOC strength,
the system favors an exotic topological structure consisting of two
criss-crossed vortex and antivortex strings as shown in figures 2(b) and
2(c), where the vortex string is distributed along the $x$ direction while
the antivortex string is distributed along the $y$ direction. Although there
are vortex-antivortex clusters or vortex-antivortex strings in individual
components, there is no phase defect in the total density distribution of
the system. It is indicated that the topological quantum states are novel
Anderson-Toulouse coreless vortex-antivortex cluster (string) states, and
have not been observed in previous studies. The vortices and antivortices in
different components repel each other and distribute staggeredly at
different positions due to the repulsion between the vortices or the
antivortices. For the case of initial component mixing and weak SOC, e.g. $%
\lambda _{x}=\lambda _{y}=1$, the ground state of the system is a
half-quantum vortex state with only a singly quantized vortex being
generated in the spin-down component (see figure 2(d)), which is somewhat
similar to that in figure 2(a). For strong SOC, the criss-crossed
vortex-antivortex string structures become noticeable (see figures 2(e) and
2(f)), which is similar to the case of initial phase separation and large
SOC strength.

What is happening physically is the spin-orbit interaction induces the
coupling between the atomic spin and the center-of-mass motion of the BEC.
Thus varying the SOC strength will lead to the change of the atomic spin
structure as well as the spin texture of the system. From figures 2(b),
2(c), 2(e) and 2(f), the vortex (antivortex) number in each component
evidently increases because the stronger SOC means there is a larger orbital
angular momentum input into the system, regardless of the initial state of
the system being mixed or separated. On the other hand, the interplay among
SOC, DDI and in-plane gradient magnetic field will change the symmetry of
vortex distribution in the system. As a result, for strong SOC, the
criss-crossed vortex-antivortex string, rather than a conventional
triangular vortex lattice, dominates the topological structure of the system.

\subsection{Role of DDI}

The next stage is to consider the effect of DDI on the ground-state
structure of initially immiscible BECs with fixed SOC and gradient magnetic
field. As mentioned before, we consider that the dipolar atoms are polarized
along the condensate plane, i.e. $\alpha =\pi /2$. Figure 3 shows the
density distributions and phase distributions for the ground states\ of the
system, where $\beta _{11}=\beta _{22}=800$, $\beta _{12}=1600$, $\lambda
_{x}=\lambda _{y}=2$, and $B=1.5$.

Considering figure 3(a), in the absence of DDI ($\gamma _{dd}=0$), there is
an antivortex in the spin-up component and the phase distributions of the
two components exhibit quadrupole-like profile, which indicates the ground
state of the system is a coreless half-quantum antivortex state with
quadrupole phase distribution. This feature evidently originates from the
combination of the effects of SOC and the in-plane gradient magnetic field
because previous studies show that the system of two-component BECs with
only SOC only supports two typical quantum phases: a plane-wave phase and a
stripe phase. In the presence of relatively weak DDI, e.g. $\gamma _{dd}=0.5$
and $\gamma _{dd}=0.7$, the system forms a sandwich-like structure, with the
two component densities being spatially separated in the central regions,
and a vortex is created in the center of the spin-down component (see
figures 3(b) and 3(c)).

For strong DDI, i.e. when $\gamma _{dd}=1$, the system sustains a particular
delaminated structure with a quadrupole phase profile in both components.
This means the density distributions of the two components are separated
fully and a prolate antivortex along the $x$ direction is generated in the
spin-up component (figure 3(d)). When DDI increases further and exceeds a
critical value, such as $\gamma _{dd}=1.2$, the system collapses and the
BECs disappear. This is due to the strong effective attraction caused by the
DDI. For the case of initial phase mixing, our simulation shows that the
ground-state structures are similar to those in the case of initial phase
separation, so there is no need to show them.

\begin{figure*}[tbp]
\centerline{\includegraphics*[width=8cm]{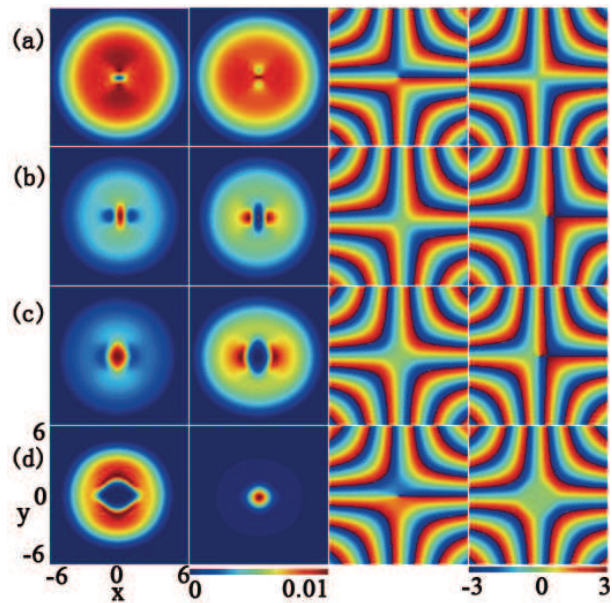}}
\caption{(Color online) Density distributions and phase distributions for
the ground states of dipolar BECs with SOC in a harmonic trap and an
in-plane gradient magnetic field, where $\protect\lambda _{x}=\protect%
\lambda _{y}=2$, $B=1.5$, $\protect\beta _{11}=\protect\beta _{22}=800$, and
$\protect\beta _{12}=1600$. The columns from left to right denote $%
\left\vert \protect\psi _{1}\right\vert ^{2}$, $\left\vert \protect\psi %
_{2}\right\vert ^{2}$, arg$\protect\psi _{1}$, and arg$\protect\psi _{2}$.
These plots are for different strengths of DDI, with (a) $\protect\gamma %
_{dd}=0$, (b) $\protect\gamma _{dd}=0.5$, (c) $\protect\gamma _{dd}=0.7$,
and (d) $\protect\gamma _{dd}=1$. The unit length is $a_{0}$.}
\label{Figure3}
\end{figure*}

\newpage To summarize the ground state structures we present table 1 where
we briefly summarize the ground-state phases and the relevant critical
values of driving parameters for different phases while maintaining $\beta
_{11}\beta _{22}<\beta _{12}^{2}$. For fixed SOC and DDI with changing
in-plane quadrupole field, as shown in figure 1, the system sustains
semicircle and crescent phase with hidden vortices, quadrupole stripe phase
with a half-quantum vortex, mixing phase with a half-quantum vortex and a
dark soliton, and miscible phase with hidden vortex-antivortex cluster,
depending on the quadrupole field strength. For fixed DDI and quadrupole
field, with the increase of SOC strength, one can observe the
Anderson-Toulouse coreless antivortex state and the criss-crossed
vortex-antivortex cluster (or string) state. Lastly we investigated the
effect of changing DDI in the presence of fixed SOC and gradient magnetic
field, as shown in figure 3, and we find that the system can exhibit a
coreless half-quantum antivortex state with quadrupole phase profile, a
sandwich structure with a half-quantum vortex, or delaminated structure with
a half-quantum antivortex.

\begin{table}[tbh]
\caption{Combined effects of SOC, DDI and quadrupole field on ground state
for $\protect\beta _{11}\protect\beta _{22}<\protect\beta _{12}^{2}$}
\label{table1}%
\begin{tabular}{cccc}
\hline\hline
SOC & DDI & Quadrupole field & Ground state \\ \hline
$\lambda _{x}=\lambda _{y}=2$ & $\gamma _{dd}=0.5$ & $B\leqslant 0.5$ &
Semicircle and crescent phase with respective hidden vortices \\
$\lambda _{x}=\lambda _{y}=2$ & $\gamma _{dd}=0.5$ & $0.5<B<1.8$ &
Quadrupole stripe phase with a half-quantum vortex \\
$\lambda _{x}=\lambda _{y}=2$ & $\gamma _{dd}=0.5$ & $1.8\leqslant B<4.2$ &
Miscible phase with a half-quantum vortex and a dark soliton \\
$\lambda _{x}=\lambda _{y}=2$ & $\gamma _{dd}=0.5$ & $B\geqslant 4.2$ &
Miscible phase with hidden vortex-antivortex cluster \\
$\lambda _{x}=\lambda _{y}<2.6$ & $\gamma _{dd}=0.5$ & $B=2.5$ &
Anderson-Toulouse coreless antivortex \\
$\lambda _{x}=\lambda _{y}\geqslant 2.6$ & $\gamma _{dd}=0.5$ & $B=2.5$ &
Criss-crossed vortex-antivortex cluster (or string) state \\
$\lambda _{x}=\lambda _{y}=2$ & $0\leqslant \gamma _{dd}<0.3$ & $B=1.5$ &
Coreless half-quantum antivortex state with quadrupole phase profile \\
$\lambda _{x}=\lambda _{y}=2$ & $0.3\leqslant \gamma _{dd}<0.9$ & $B=1.5$ &
Sandwich structure with a half-quantum vortex \\
$\lambda _{x}=\lambda _{y}=2$ & $0.9\leqslant \gamma _{dd}<1.2$ & $B=1.5$ &
Delaminated structure with a half-quantum antivortex \\
$\lambda _{x}=\lambda _{y}=2$ & $\gamma _{dd}\geqslant 1.2$ & $B=1.5$ &
System collapses \\ \hline\hline
\end{tabular}%
\end{table}

\subsection{Spin textures}

Now we analyze the spin densities and spin textures of the system in order
to further elucidate the ground-state properties. To consider this one looks
at figure 4, which show the representative spin-density distributions (the
top three rows) and the corresponding topological charge densities (the
bottom row). The relevant parameters for the four cases are: (a) $\beta
_{11}=\beta _{22}=800$, $\beta _{12}=1600$, $\lambda _{x}=\lambda _{y}=2$, $%
\gamma _{dd}=0.5$, $B=6$; (b) $\beta _{11}=\beta _{22}=800$, $\beta
_{12}=1600$, $\lambda _{x}=\lambda _{y}=3$, $\gamma _{dd}=0.5$, $B=2.5$; (c)
$\beta _{11}=\beta _{22}=800$, $\beta _{12}=600$, $\lambda _{x}=\lambda
_{y}=5$, $\gamma _{dd}=0.5$, $B=2.5$ and (d) $\beta _{11}=\beta _{22}=800$, $%
\beta _{12}=1600$, $\lambda _{x}=\lambda _{y}=2$, $\gamma _{dd}=1$, $B=1.5$.
The density distributions and phase distributions corresponding to figures
4(a)-4(d) are given in figure 1(d), figure 2(b), figure 2(e), and figure
3(d), respectively. In the pseudo-spin representation, the red region
denotes spin-up and the blue region denotes spin-down. From figure 4, $S_{x}$
obeys an odd-parity distribution along the $x$ direction and an even-parity
distribution along the $y$ direction, while the situation is the reverse for
$S_{y}$. At the same time, $S_{z}$ and $q(r)$ satisfy an even-parity
distribution along both the $x$ direction and the $y$ direction.

\begin{figure*}[tbp]
\centerline{\includegraphics*[width=11cm]{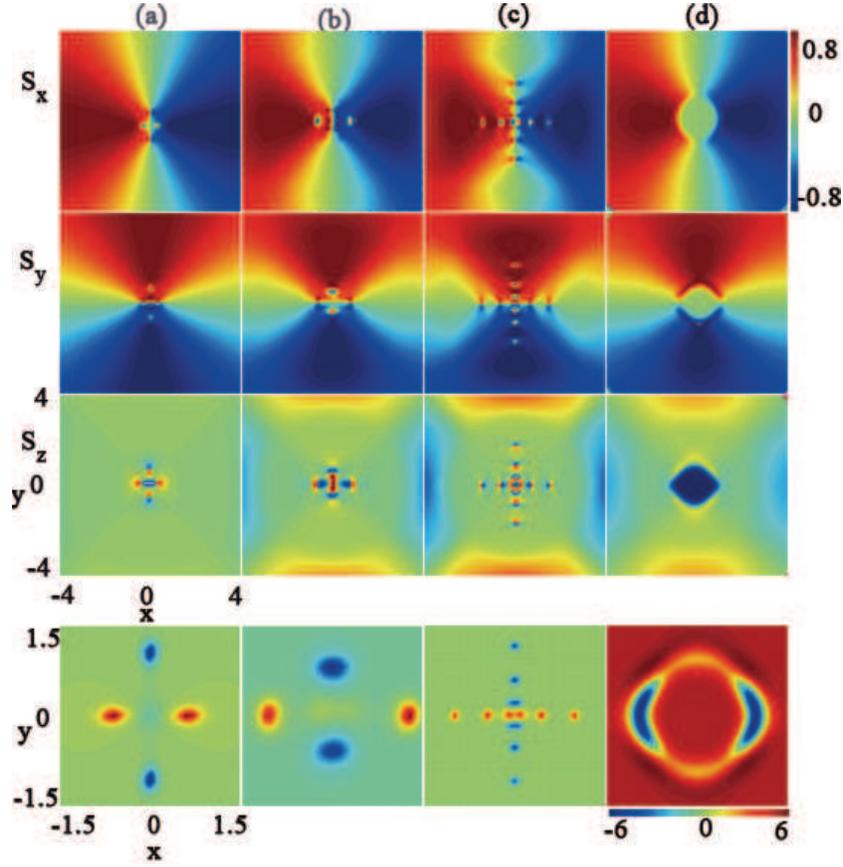}}
\caption{(Color online) Spin densities and topological charge densities of
spin-orbit-coupled dipolar BECs in a harmonic trap and an in-plane
quadrupole field. (a) $\protect\beta _{11}=\protect\beta _{22}=800$, $%
\protect\beta _{12}=1600$, $\protect\lambda _{x}=\protect\lambda _{y}=2$, $%
\protect\gamma _{dd}=0.5$, $B=6$, (b) $\protect\beta _{11}=\protect\beta %
_{22}=800$, $\protect\beta _{12}=1600$, $\protect\lambda _{x}=\protect%
\lambda _{y}=3$, $\protect\gamma _{dd}=0.5$, $B=2.5$, (c) $\protect\beta %
_{11}=\protect\beta _{22}=800$, $\protect\beta _{12}=600$, $\protect\lambda %
_{x}=\protect\lambda _{y}=5$, $\protect\gamma _{dd}=0.5$, $B=2.5$, and (d) $%
\protect\beta _{11}=\protect\beta _{22}=800$, $\protect\beta _{12}=1600$, $%
\protect\lambda _{x}=\protect\lambda _{y}=2$, $\protect\gamma _{dd}=1$, $%
B=1.5$. The rows (from top to bottom) denote $S_{x}$, $S_{y}$ and $S_{z}$
components of the spin density vector, and the topological charge density $%
q(r)$. The density distributions and phase distributions corresponding to
(a)-(d) are given in Fig. 1(d), Fig. 2(b), Fig. 2(e), and Fig. 3(d),
respectively. The unit length is $a_{0}$.}
\label{Figure4}
\end{figure*}

In addition, there are criss-cross crystal-like structures in the $S_{x}$, $%
S_{y}$, $S_{z}$ components of the spin density and the topological charge
density $q(r)$ (see figures 4(a)-4(c)), which indicates that there exist
special spin defects in the present system. Furthermore, two remarkable spin
domains are generated in spin components $S_{x}$ and $S_{y}$, and the
boundary between the two spin domains forms a peculiar hyperbolic spin
domain wall with $\left\vert S_{x}\right\vert \neq 1$ and $\left\vert
S_{y}\right\vert \neq 1$ due to the presence of in-plane quadrupole filed,
which can be seen clearly in the top two rows of figure 4. It is well known
that the spin domain wall for a system consisting of two-component BECs is
typically a classical Neel wall, where the spin flips only along the
perpendicular direction of the wall. However, our simulation shows that in
the region of the domain wall the spin flips not only along the vertical
direction of domain wall, but also along the domain-wall direction, which
indicates that the observed hyperbolic domain wall is a new type of domain
wall.

In figure 5, we give the spin textures of the system, where the parameters
for the four rows are the same with those for figures 4(a)-4(d). Shown in
the right three columns are the local enlargements of the spin textures in
the left first column, so one can better see the details of the spin
textures. Our simulation shows that the spin textures in figure 5 exhibit
strong symmetry with respect to the $y=0$ axis and the $x=0$ axis, where the
texture singularities are symmetrically distributed along the two axes. The
excellent symmetry is also reflected in the topological charge densities in
the bottom row of figure 4. In the context of this work, one only needs to
focus on the left region, the upper region and the central region of the
spin texture, as shown in columns 2-4 (from left to right) of figure 5. Our
numerical calculation shows that the local topological charges in figures
5(a2), 5(a3) and 5(a4) approach $Q=1$, $Q=-1$, and $Q=-0.5$, respectively,
which indicates the local topological defects in figures 5(a2), 5(a3) and
5(a4) are hyperbolic-radial(out) skyrmion, hyperbolic-radial(in)
antiskyrmion, and hyperbolic half-antiskyrmion (antimeron) \cite%
{Su,Shi,CLiu,Mermin}, respectively. Thus the texture in figure 5(a1) forms a
special skyrmion-half-antiskyrmion-antiskyrmion lattice composed of two
hyperbolic-radial(out) skyrmions along the $x$ direction, two
hyperbolic-radial(in) skyrmions along the $y$ direction, and a hyperbolic
half-antiskyrmion in the center.

\begin{figure*}[tbp]
\centerline{\includegraphics*[width=13cm]{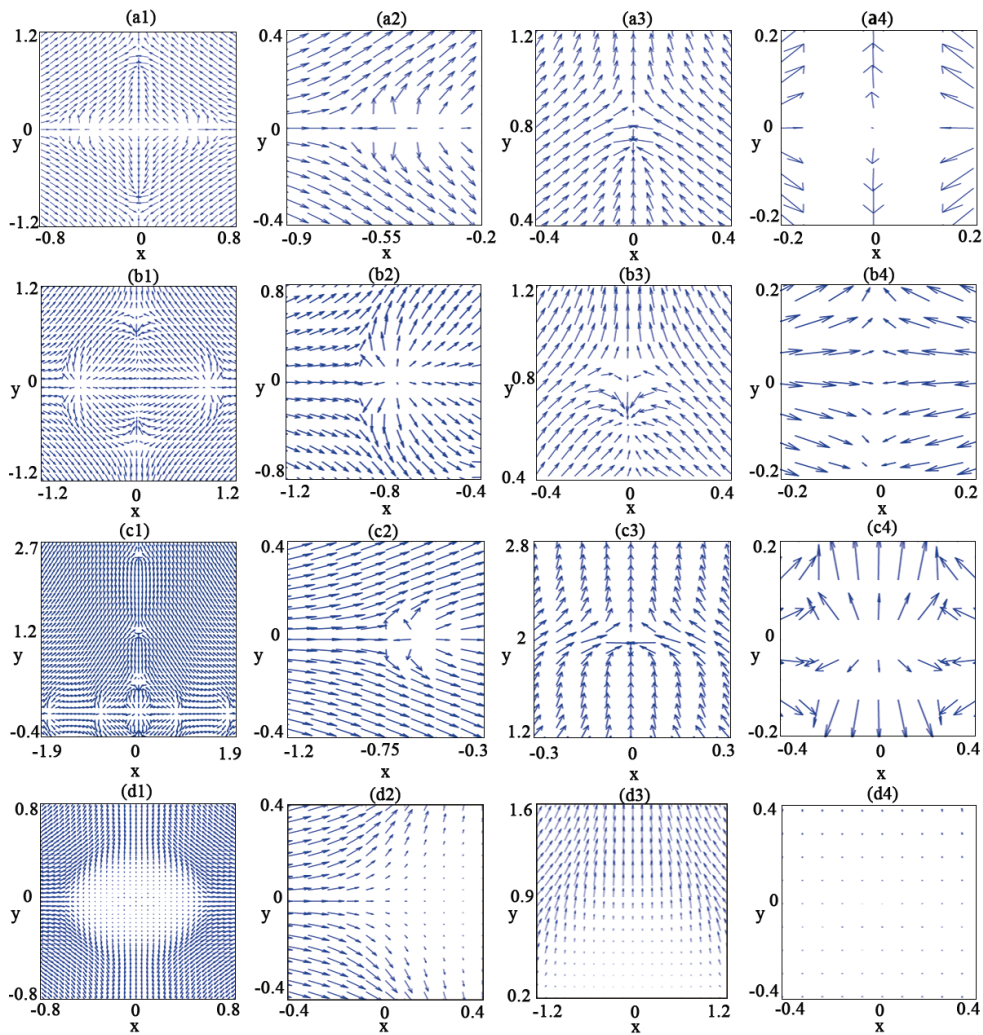}}
\caption{(Color online) Spin textures of dipolar BECs with SOC in a harmonic
trap and an in-plane quadrupole field, where the corresponding spin
densities and topological charge densities are given in Fig. 4. The
parameters in the four rows from top to bottom are the same with those in
Figs. 4(a)-4(d), respectively. Shown in left column 1 are the spin textures
for various parameter values, and displayed in the right three columns are
the local enlargements of the spin textures in the left first column. The
unit length is $a_{0}$.}
\label{Figure5}
\end{figure*}

Though the component density distributions for the ground state in figure
2(b) are quite different from those in figure 1(d), the spin texture in
figure 5(b1) is somewhat similar to that in figure 5(a1) and also posses
strong symmetry. The local topological charge for the left spin defect (see
figure 5(b2)) and the right spin defect is $Q=1$; and that for the upper
spin defect (see figure 5(b3)) and the lower spin defect is $Q=-1$. The
difference is that the local topological charge of the central spin defect
(see figure 5(b4)) is $Q=0.5$. That is, the topological structure in figure
5(b1) is a skyrmion-half-skyrmion-antiskyrmion lattice made of
hyperbolic-radial(out) skyrmions, hyperbolic-radial(in) skyrmions and a
hyperbolic half-skyrmion (meron).

The spin texture for the ground state shown in figure 2(e) is richer and
more interesting, which has SOC, DDI, quadrupole field and there is initial
component mixing. Considering figure 5(c1), we see that it has good symmetry
but also posses complexity in the spin texture. For the convenience of
analysis, we mainly show the region of $y>0$ of the spin texture, where the
representative local enlargements are displayed in figures 5(c2)-5(c4). Our
calculation results show that the topological charge for individual spin
defects (except for the central spin defect) along the $x$ the $y$ axes are $%
Q=0.5$ and $Q=-0.5$, respectively. This implies that the horizontal
topological defects (except for the central defect) are hyperbolic
half-skyrmions (merons) (see figure 5(c2)) and the vertical ones (except for
the central defect) are hyperbolic half-antiskyrmions (antimerons) (see
figure 5(c3)). The central topological defect is a radial-out skyrmion with
local topological charge $Q=1$ (see figure 5(c4)). Therefore the topological
structure in figure 5(c1) is an exotic
half-skyrmion-skyrmion-half-antiskyrmion lattice composed of four horizontal
half-skyrmions, one central skyrmion and six vertical half-antiskyrmions.
Displayed in figure 5(d1) is a half-antiskyrmion (antimeron) with
topological charge $Q=-0.5$. Compared with conventional half-skyrmion, there
exists a large singularity region with a waist-drum shape in the spin
texture (see figures 5(d1)-5(d4)), which is essentially caused by the
presence of in-plane quadrupole field and strong DDI. To the best of our
knowledge, these new topological structures mentioned above have not\ been
reported in previous studies, and can be observed in future experiments.


\section{Conclusion}

We have studied a rich variety of ground-state phases and topological
defects of quasi-2D two-component dipolar BECs with Rashba SOC in a harmonic
trap and an in-plane gradient magnetic field. The combined effects of the
in-plane gradient magnetic field, SOC, DDI, and interatomic interactions on
the ground-state structure of this system were discussed systematically. Our
results show that for strong quadrupole field and fixed SOC and DDI
strengths the system favors a ring mixed phase with a hidden
vortex-antivortex lattice (cluster) rather than a usual giant vortex in each
component. In particular, for the case of initial component separation and
relatively weak quadrupole field, the ground state is an unusual quadrupole
stripe phase with a half-quantum vortex or a quadrupole TF phase with a
half-quantum antivortex. For given DDI and quadrupole field strengths,
increasing SOC strength leads to the formation of a criss-crossed
vortex-antivortex string state. When the strengths of SOC and quadrupole
field are fixed, the ground state for strong DDI exhibits a sandwich-like
structure, or a special delaminated structure with a prolate antivortex in
the spin-up component and quadrupole phase distribution in both components.
The results for different parameter regimes can be easily seen in table 1.
Furthermore, the typical spin textures of the system were analyzed. We find
that the system supports exotic new topological defects, such as a
hyperbolic spin domain wall, skyrmion-half-antiskyrmion(or
half--skyrmion)-antiskyrmion lattice,
half-skyrmion-skyrmion-half-antiskyrmion lattice, and drum-shaped
half-antiskyrmion (antimeron). This work has numerically demonstrated the
many complex, novel and interesting topological structures that are present
in this system, and since it is expected that these results are to be
verified in future experiments, this paper could serve as a valuable
resource for experimentalists working on these systems and comparing theory
and experiment. These findings have enriched our new understanding for
topological excitations in ultracold atomic gases and condensed matter
physics.

\begin{acknowledgments}
L.W. thanks Professor Yongping Zhang and Professor Yong Xu for their helpful
discussions, and acknowledges the research group of Professor W. Vincent Liu
at the University of Pittsburgh, where part of the work was carried out.
This work was supported by the National Natural Science Foundation of China
(Grant Nos. 11475144 and 11047033), the Natural Science Foundation of Hebei
Province of China (Grant Nos. A2019203049 and A2015203037), and Research
Foundation of Yanshan University (Grant No. B846).
\end{acknowledgments}

\end{document}